# RAPID CAVITY PROTOTYPING USING MODE MATCHING AND GLOBALISED SCATTERING MATRIX

I. Shinton and R.M. Jones; Cockcroft institute, Daresbury; and The University of Manchester, UK

*Abstract*

Cavity design using traditional mesh based numerical means (such as the finite element or finite difference methods) require large mesh calculations in order to obtain accurate values and cavity optimisation is often not achieved. Here we present a mode matching scheme which utilises a globalised scattering matrix approach that allows cavities with curved surfaces (i.e. cavities with elliptical iris's and or equators) to be accurately simulated allowing rapid cavity prototyping and optimisation to be achieved. Results on structures in the CLIC main linacs are presented.

## INTRODUCTION

The mode matching method is a relatively mature electromagnetic concept in which the analytical solutions of Maxwell's equations are given as a series expansion of modes; the electromagnetic fields are subsequently obtained by field matching at the interfaces between the sub-regions of a given structure [1] - [7]. The method allows rapid field calculations to be obtained and provided a correct mode ratio and filtering (to circumvent the overshoot due to the Gibbs phenomena of the field at the junction) are both implemented [3] an exact solution of Maxwell's equations will be yielded.

## RAPID CAVITY DESIGN USING AN ANALYTICAL DISPERSION EQUATION

In [8] a mode matching method was combined with a GSM technique, which gave rise to an analytical dispersion relationship given by Eq. 1.

$$\cos\Phi = \frac{1+S_{21}(1,1)^2 - S_{11}(1,1)^2}{2S_{21}(1,1)} \qquad (1)$$

Eq. 1 was obtained by assuming an infinitely periodic structure for a two port system, in which there is only one propagating mode within the transition and secondly that the evanescent modes have sufficiently decayed from the transition. In Eq. 1 $\Phi$ is the phase advance per cell and $S$ is the Scattering matrix in which the subscripts represent the waves scattered from the ports.

The S matrix at the transition is obtained by mode matching the fields in terms of the modal amplitudes, in which we consider the waves to be propagating from one port to the other [9]. All the S matrices presented were calculated semi-analytically in terms of a series of WN (wide narrow) or NW (narrow wide) transitions using the methodology and equations presented in [8]. This mitigates the need to numerically determine the scattering matrix for a structure. The following terminology will be used to describe the geometry of cavities within this paper: iris radius "a", equator radius "b", cell length "L", iris thickness "t" and the iris ellipticity is describe by the vertical axis "a1" and horizontal axis "a2". Finally the term WNW is used to describe a wide narrow wide sharp transition.

In this paper the numerical comparisons were performed using the eigen solver of HFSSv11; convergence studies were preformed and the points presented on all the curves represent the extrapolations for an infinite mesh. These results agree with those presented in [8] which were calculated using HFSSv8.5, however we note that the results using version 11 are more accurate.

## ANALYTICAL DISPERSION CURVES FOR SHARP TRANSITIONS

Below in Fig. 1 is a comparison of the dispersion curves obtained using Eq. 1, against the numerical results of HFSS and those obtained using KN7c [10]. The maximum error between individual methods is less than 2MHz.

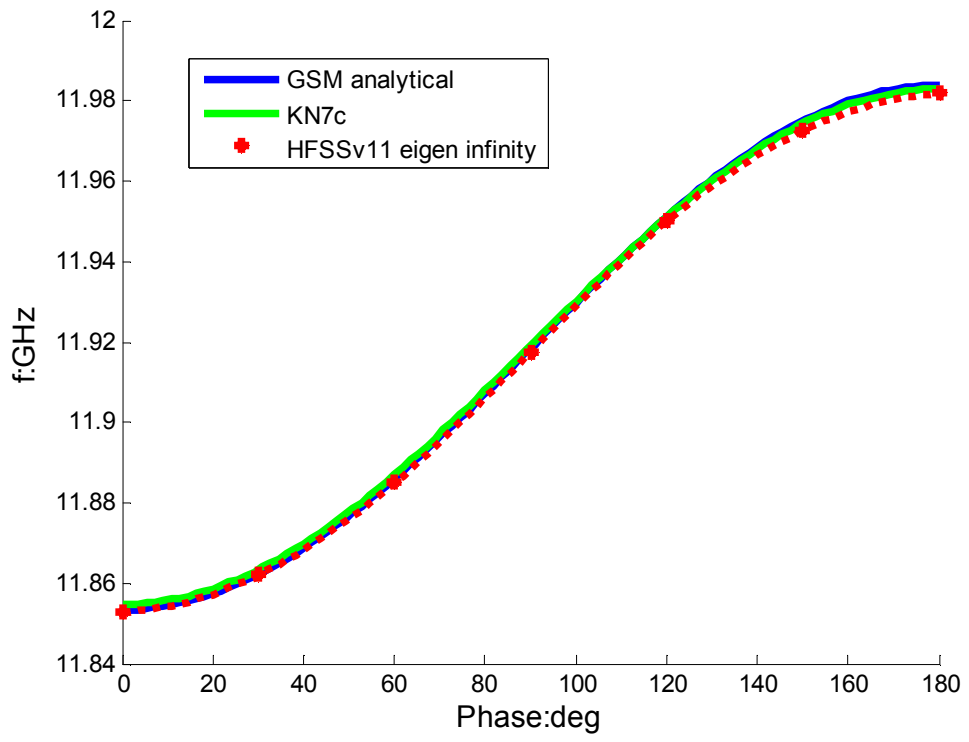

Figure 1: Comparison of dispersion curves calculated using analytical (Eq. 1 blue) and numerical approaches (HFSSv11 extrapolated to infinite mesh red and KN7c green) for an idealised CLIC G cell [11], which has been represented as a simplified WNW transition.

Provided the number of modes in the iris "a" is less than or equal to the modes in the equator "b" the solution will be unique [12], [13]. Studies have been preformed to determine the optimal ratio of modes to allow for convergence [14]; however in practice the ratio of the equator to iris radii is very close to that of the optimal mode ratio, refer to [15] in which a proof is given for the limit of this ratio. Fig. 2 presents this between the convergence of an equal mode number to that for an optimal mode ratio, where we observe that infinite convergence is reached much faster using an optimal mode ratio.

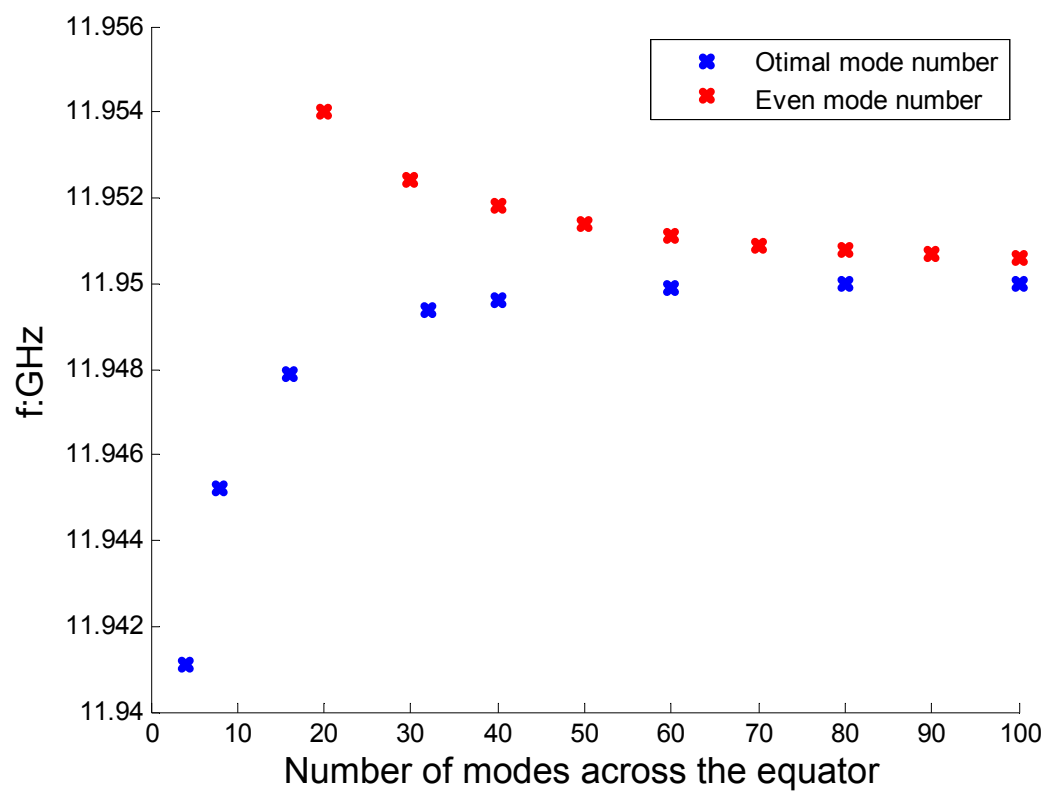

Figure 2: Convergence study of a WNW transition of dimensions a=2.69mm, b=9.78mm, t=1.2mm and L=8.33mm – here we see that using an optimised mode ratio (3:1) infinite convergence is reach just after 60 modes.

There are two distinct advantages of the analytical approach presented in this paper, the first is the amount of computational time (and resources) saved as compared to a purely numerical approach. The analytical dispersion curve predicted using Eq. 1 is obtained as a single calculation taking less than a minute, whereas each point obtained using HFSS required approximately 45mins. Secondly provided a sufficiently large number of modes are used in a mode matching scheme then the necessity to carry out a convergence study is negated.

## ANALYTICAL DISPERISION CURVES FOR ELLIPTICAL AND CURVED CAVITY STRUCTURES

A few early mode matching schemes relied upon approximating the curved iris in X-band structures by averaging the iris dimensions in terms of a single sharp transition. This is unsatisfactory as pointed out in various papers [16], for although generic design factors can be obtained it is a somewhat an ad hoc procedure.

The methodology presented in the previous section can be extended in which a curved surface is represented as a series of conjoined WN and or NW transitions. Displayed in Fig 3 is a comparison of the dispersion curves obtained for the first CLIC ZC cell [17] (with a curved iris) using HFSSv11 and the aforementioned method. A total of 40 evenly spaced segments in the iris, and an even number of 60 modes across each transition were used in the mode matching method that utilises Eq. 1. The results are in very good agreement and vary less than a maximum difference of 3MHz.The process can be thought of as being similar to how basic finite difference schemes approximate curved surfaces. The greater the number of segments used in the method the more accurate the answer; however a convergence plateau is reached (as can be observed in Fig 4) after which numerical round-off starts to accumulate.

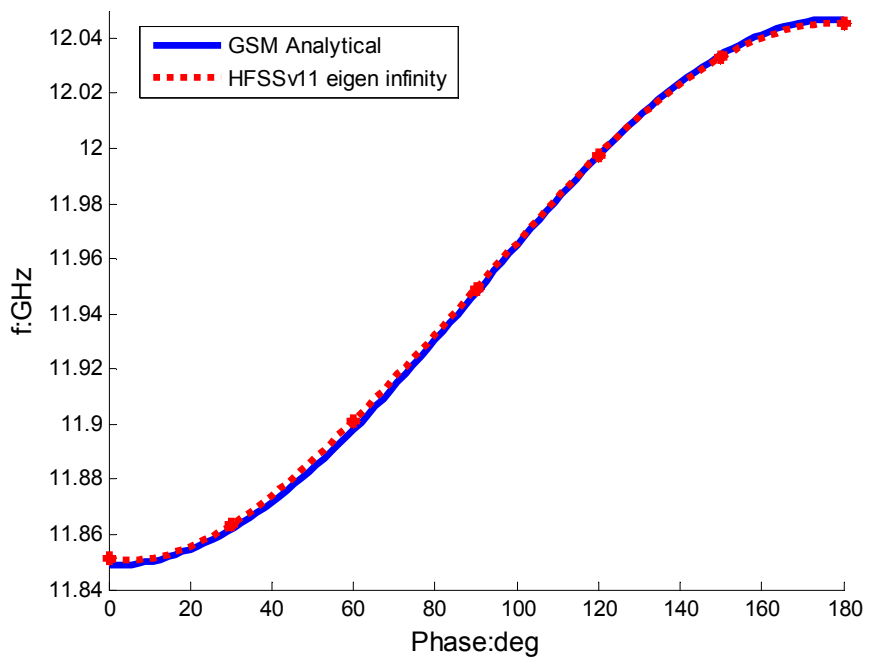

Figure 3: Dispersion curves calculated using analytical (Eq. 1 blue) and numerically using HFSSv11 (extrapolated to infinite mesh) for the first CLIC ZC cell. Dimensions a=2.99mm, b=9.88mm, t=0.8mm, a1=0.45mm and a2=t.

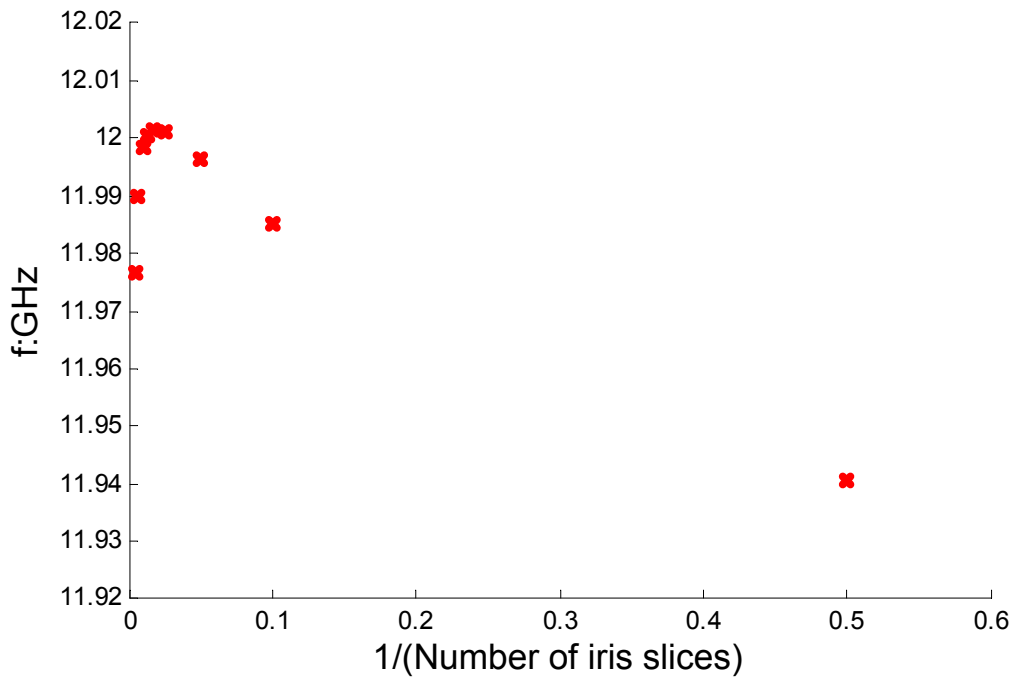

Figure 4: Convergence study of the CLIC ZC cell 1 using the GSM mode matching technique, where the curved surface is approximated by a series of WN and NW transitions.

## RAPID CAVITY OPTIMISATION

Since the method allows an analytical dispersion curve to be obtained as a single calculation, calculations of the electromagnetic fields within the cell are performed concurrently [8]. This allows investigation of the parameter space by varying several aspects of the geometry of the cell in question iteratively and recording the effect on the desired design parameters. As an example of the accuracy of this technique a comparison of the R/Q calculated using HFSS and the GSM mode matching method for a sharp WNW transition yields a percentage difference of 1.12% (dimensions of a=2.99mm, b=9.88mm, t=0.8mm and L=8.33mm at a phase advance of 2/3π).

In order to obtain the most advantageous design at a designated phase advance, the various geometric parameters are iteratively varied. Displayed below in Fig 5 is an example of such a parameter sweep, in which we have defined a 2/3π phase advance and calculate the corresponding effect on two of the design parameters. We apply this technique to the potential CLIC structure known as CLIC-ZC [17]. The frequency dependence on iris thickness and radius ellipticity is illustrated in Fig. 6.

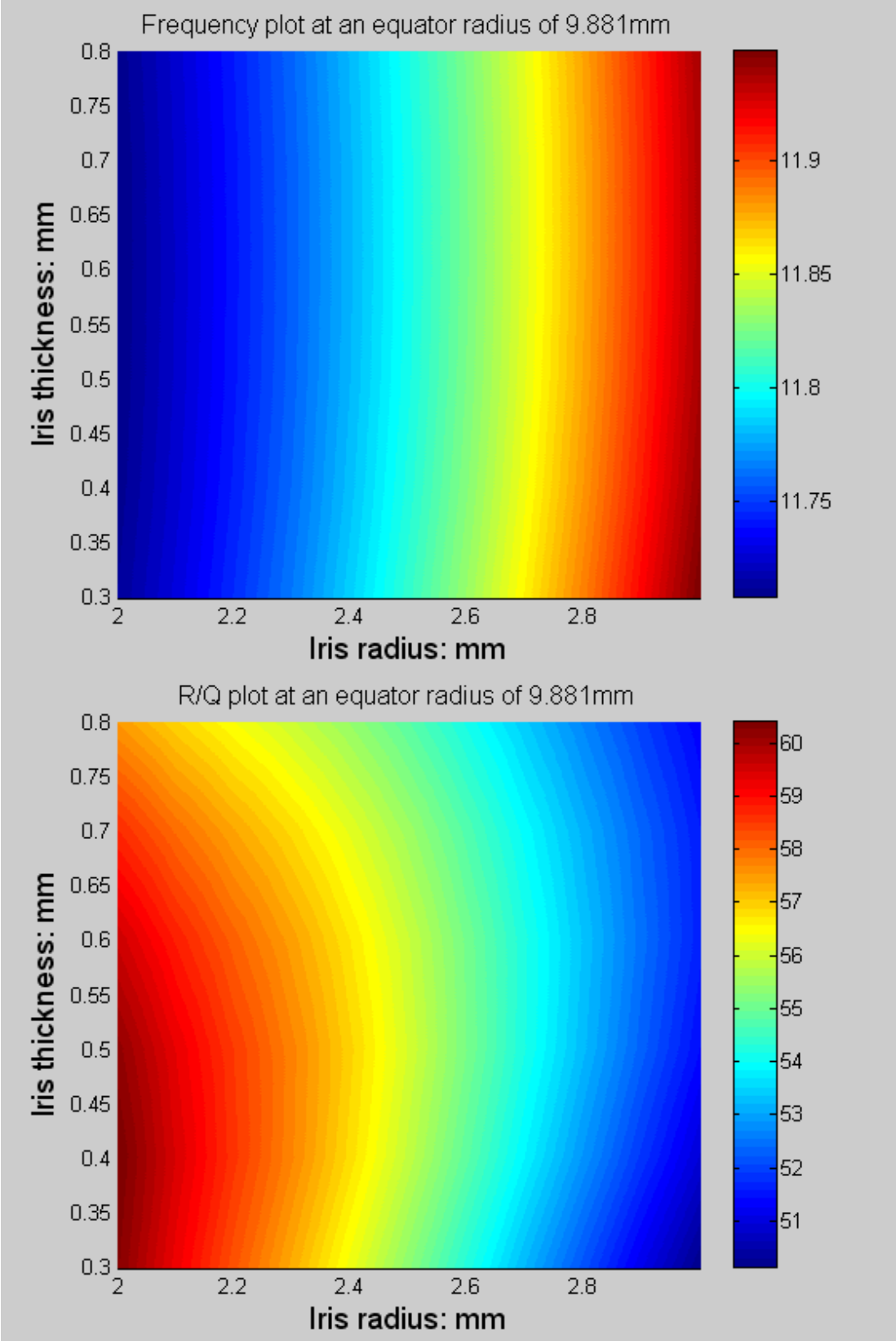

Figure 5: Two slices of the iterative dimensional results for a WNW transition for the variation of iris thickness and iris radius for a particular equator radius b=9.88mm at $2/3\pi$ phase advance. The colour axis represents the frequency in GHz and R/Q in Ohms respectively in the first and second plots.

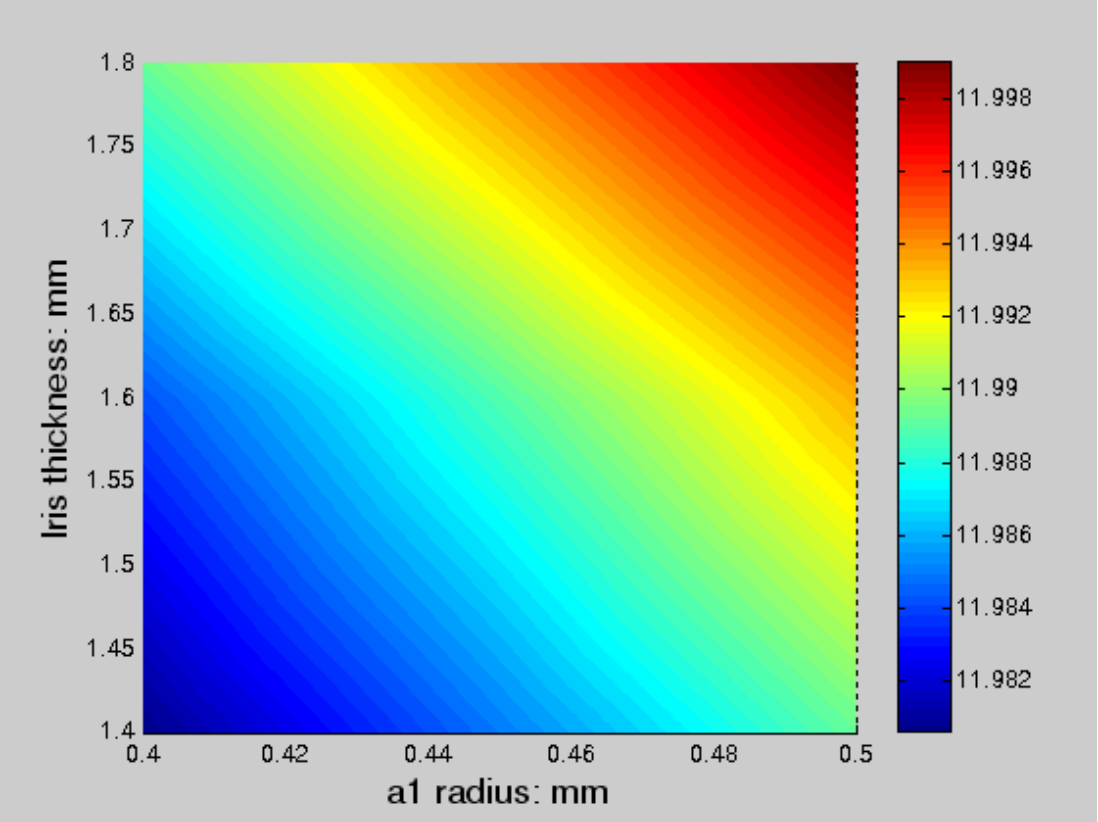

Figure 6: A slice of the results for a CLIC-ZC cell for the variation of iris thickness and the vertical iris ellipse radius a1 for a particular equator radius b=9.881mm at a $2/3\pi$ phase advance. The colour axis represents the frequency in GHz.

## DISCUSSION

Mode matching is a straightforward powerful cavity design tool that can be readily employed to model cavities as it is rapid and accurate. Curved geometrise are readily characterised using multiply connected sharp transitions. However this method suffers from numerical round-off errors. Using tapered transitions [16] rather than sharp transitions would improve the representation of curved geometries. The method could be further improved firstly by incorporating optimisation routines, for example the simplex method.